\documentclass[sigconf,authorversion]{acmart}

\AtBeginDocument{%
  \providecommand\BibTeX{{%
    \normalfont B\kern-0.5em{\scshape i\kern-0.25em b}\kern-0.8em\TeX}}}

\copyrightyear{2025} 
\acmYear{2025} 
\setcopyright{acmlicensed}\acmConference[WSDM '25]{Proceedings of the Eighteenth ACM International Conference on Web Search and Data Mining}{March 10--14, 2025}{Hannover, Germany}
\acmBooktitle{Proceedings of the Eighteenth ACM International Conference on Web Search and Data Mining (WSDM '25), March 10--14, 2025, Hannover, Germany}
\acmDOI{10.1145/3701551.3704120}
\acmISBN{979-8-4007-1329-3/25/03}

\usepackage{booktabs} 
\usepackage{color}
\usepackage{makecell}
\usepackage{subcaption}
\usepackage{multirow}
\usepackage{enumitem}

\newcommand{\shrink}{\vspace*{-.9\baselineskip}}

\begin{document}

\fancyhead{}
\title{CRS Arena: Crowdsourced Benchmarking of Conversational Recommender Systems} 

\author{Nolwenn Bernard}
\affiliation{%
  \institution{University of Stavanger}
  \city{Stavanger}
  \country{Norway}
}
\email{nolwenn.m.bernard@uis.no}

\author{Hideaki Joko}
\affiliation{%
  \institution{Radboud University}
  \city{Nijmegen}
  \country{The Netherlands}
}
\email{hideaki.joko@ru.nl}

\author{Faegheh Hasibi}
\affiliation{%
  \institution{Radboud University}
  \city{Nijmegen}
  \country{The Netherlands}
}
\email{faegheh.hasibi@ru.nl}

\author{Krisztian Balog}
\affiliation{%
  \institution{University of Stavanger}
  \city{Stavanger}
  \country{Norway}
}
\email{krisztian.balog@uis.no}

\begin{abstract}
We introduce \emph{CRS Arena}, a research platform for scalable benchmarking of Conversational Recommender Systems (CRS) based on human feedback. %
The platform displays pairwise battles between anonymous conversational recommender systems, where users interact with the systems one after the other before declaring either a winner or a draw. CRS Arena collects conversations and user feedback, providing a foundation for reliable evaluation and ranking of CRSs. %
We conduct experiments with CRS Arena on both open and closed crowdsourcing platforms, confirming that both setups produce highly correlated rankings of CRSs and conversations with similar characteristics. We release \emph{CRSArena-Dial}, a dataset of 474 conversations and their corresponding user feedback, along with a preliminary ranking of the systems based on the Elo rating system. 
The platform is accessible at \url{https://iai-group-crsarena.hf.space/}.
\end{abstract}




\maketitle

\section{Introduction}

Conversational recommender systems (CRSs) are attracting attention due to their ability to provide personalized recommendations. Indeed, unlike traditional recommender systems, CRSs are interactive, offering users the possibility to express their current preferences and provide direct feedback on recommended items through natural language~\citep{Jannach:2021:CSUR}. Despite the potential of CRSs, current evaluation often relies on offline metrics or user studies that focus on some specific aspect of the system, which is assessed based on static dialogue corpora. For example, \citet{Manzoor:2021:RecSys} focus only on the quality of response generation in pre-defined context, i.e., do not consider a sequence of generated responses. 
Therefore, we argue that current evaluations often overlook the interactive nature of the problem.
This may be due to the fact that most CRSs are research prototypes and performing a fair comparison of them with real users is both time consuming and expensive.
While user simulation can alleviate some of these issues~\citep{Balog:2024:FnTIR}, simulation-based results are only indicative and need to be validated with real users.

In this work, we introduce CRS Arena, a platform to benchmark existing CRSs in a crowdsourced environment that is inspired by Chatbot Arena~\citep{Chiang:2024:ICML}.
In CRS Arena, pairs of CRSs face each other in side-by-side ``battles'' where users can interact with them and declare a winner or a tie. 
The platform makes the evaluation process and CRSs accessible to a wide range of users, including those without technical expertise, in a fun and engaging way.
Based on the outcomes of battles, a ranking of CRSs can be established based on first-party user feedback. In addition to the pairwise preferences (i.e., battle outcomes), users also leave explicit feedback regarding their (dis)satisfaction with the individual CRSs. 

We tested CRS Arena in two crowdsourced environments, open and closed. Open corresponds to public access, while closed means restricted access to a selected group of workers recruited using a crowdsourcing platform. 
Using nine CRSs, we collected a total of 474 conversations, annotated with human feedback regarding overall satisfaction and battle outcome. 
The resulting dataset, \emph{CRSArena-Dial}, is unique in that it contains conversations between CRSs and real users as opposed to a Wizard-of-Oz setting that is typically employed in existing dialogue corpora (i.e., there is a human worker acting as the CRS)~\citep{Li:2018:NIPS,Joko:2024:SIGIR}.
The dataset is made available to the community for further research and can be analyzed to better understand the capabilities and limitations of CRSs. 
Additionally, we compute Elo rating for each CRS based on the results of the battles to rank them. 
We find that the Elo ranking disagrees with the recall-based ranking reported for these systems, and user feedback shows overall low user satisfaction, highlighting the importance of considering actual users and the interactive nature of the task when evaluating CRSs.

In summary, the main contributions of this work are:
\begin{itemize}
    \item We develop CRS Arena, a platform where users can interact with CRSs and evaluate them in a realistic setting. 
    \item We release a dataset of 474 conversations with various CRSs, along with first-party user feedback, collected using CRS Arena in both open and closed crowdsourced environments. 
    \item We provide an initial analysis of the collected data, examining conversation characteristics, human feedback, and ranking of CRSs based on Elo rating.
    \item We demonstrate the robustness of CRS Arena across various crowdsourcing setups, showing its feasibility for fast and scalable human evaluation of CRSs.
\end{itemize}
CRS Arena is open source and accessible at \url{https://iai-group-crsarena.hf.space/}.\footnote{The Safari browser is currently not supported.}

\vspace*{-.6\baselineskip}
\section{Related work}
\label{sec:related}

The evaluation of conversational recommender systems (CRSs) has been identified as an open challenge in the field~\citep{Gao:2021:AIOpen}. Indeed, it is a complex task that requires the consideration of different aspects from the system and user perspectives, in addition to being highly interactive. Different evaluation methodologies have been proposed in the literature, including online experiments, user, and computational studies~\citep{Jannach:2021:CSUR}, with the last two being the most common.
Toolkits and frameworks have been developed to facilitate the evaluation of CRSs, including CRSLab~\citep{Zhou:2021:ACL} and iEvaLM~\citep{Wang:2023:EMNLP}, which provide resources for computational studies, and INFACT~\citep{Manzoor:2022:KARS} and CRS-Que~\citep{Jin:2024:TOIS}, which facilitate user studies.

We make the following observations based on our examination of the literature. First, there is a lack of standardization in the evaluation of CRSs, which makes their comparison cumbersome.
Second, directly interacting with CRSs is challenging due to the lack of user interface in existing toolkits and frameworks, e.g., CRSLab has an interface for only one out of five CRSs.
Third, evaluation often focuses on specific aspects of the system, such as recommendation quality or the fluency of the responses, rather than the overall user experience. Finally, user studies often use grading rubrics resulting in pointwise comparisons; due to the subjective nature of the evaluation it can be difficult to ensure the consistency of the grading across many evaluators~\citep{Kulikov:2019:ILNG}.

To mitigate some of these points, we propose CRS Arena, a platform to collect conversations and human feedback on CRSs in a realistic setting and to rank them based on pairwise comparisons.
Generally, relative (pairwise) comparisons are easier for people and yield greater agreement among assessors than pointwise comparisons~\citep{Carterette:2008:ECIR}.
Our platform is inspired by the Chatbot Arena~\citep{Chiang:2024:ICML} that facilitates the benchmarking of large language models (LLMs) based on pairwise comparisons. The platform has led to the release of valuable resources, e.g., a dataset of realistic prompts and a leaderboard, for the community. We believe that CRS Arena can have a similar impact on the benchmarking of CRSs.
The main differences between Chatbot Arena and CRS Arena are: (1) the focus, i.e., LLMs vs. CRSs, (2) the independence of the systems, i.e., in Chatbot Arena the systems receive the same prompts while in CRS Arena the conversations are independent, and (3) CRS Arena collects users' explicit feedback on task success, i.e., frustration or satisfaction, after each conversation.

\vspace*{-.6\baselineskip}
\section{CRS Arena}
\label{sec:arena}

\begin{figure}
    \shrink
    \centering
    \includegraphics[width=\columnwidth]{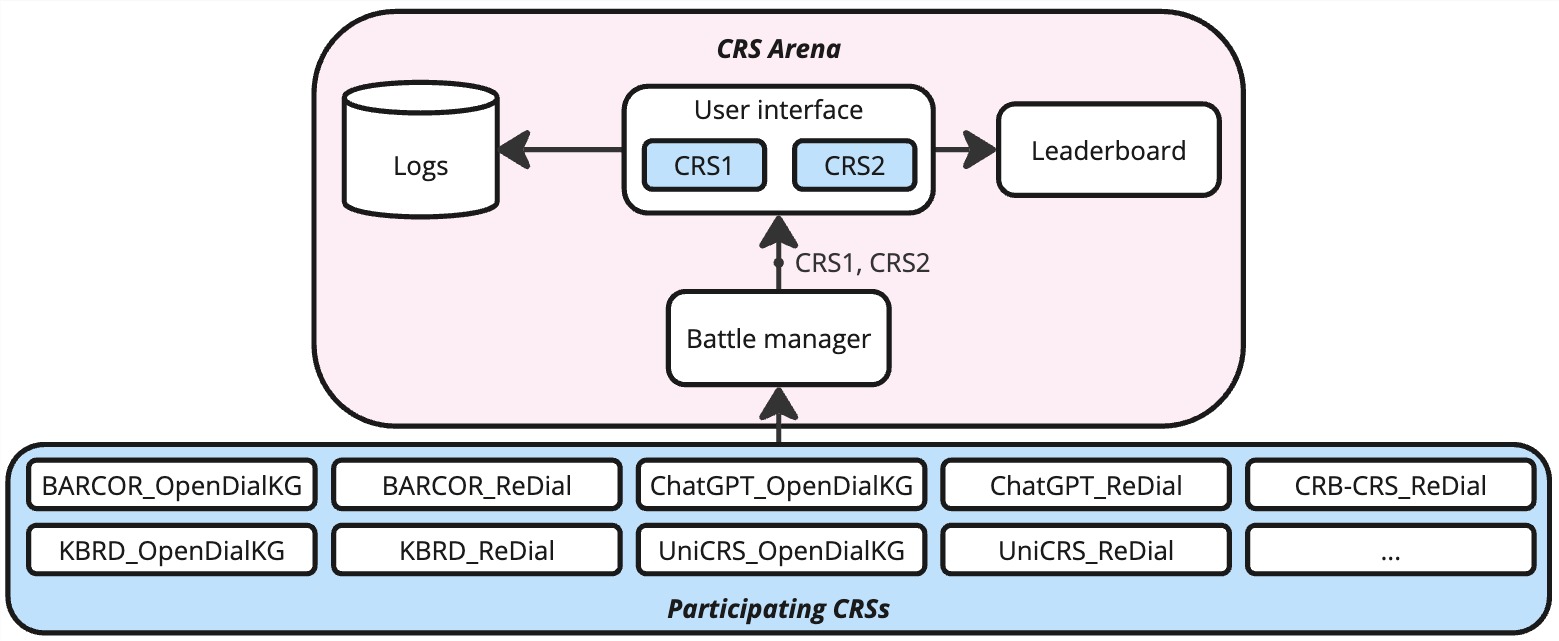}
    \shrink
    \caption{Overview of the main components of CRS Arena.}
    \label{fig:overview}
    \shrink
\end{figure}

This work presents CRS Arena, a platform for benchmarking conversational recommender systems (CRSs) in a realistic setting. The main components of the platform are shown in Fig.~\ref{fig:overview}.

\vspace*{-.6\baselineskip}
\subsection{Conversational Recommender Systems}
\label{sec:arena:crs}

Currently, there are nine CRSs available in the arena; see \emph{\textbf{Participating CRSs}} in Fig.~\ref{fig:overview}. These include \emph{KBRD}~\citep{Chen:2019:EMNLP}, \emph{BARCOR}~\citep{Wang:2022:arXiv}, \emph{UniCRS}~\citep{Wang:2022:KDD}, \emph{ChatGPT}~\citep{Wang:2023:EMNLP}, and \emph{CRB-CRS}~\citep{Manzoor:2022:InfSys}, which are trained and/or leverage external knowledge from either OpenDialKG~\citep{Moon:2019:ACL} or ReDial~\citep{Li:2018:NIPS}.
These CRSs are implemented in an extended version of the iEvaLM framework~\citep{Wang:2023:EMNLP} that facilitates the integration of new CRSs. They only need to inherit from the CRS base class and implement the method responsible for generating responses to incoming user utterances. Consequently, new participating systems can join the arena with minimal effort.

\vspace*{-.6\baselineskip}
\subsection{Battle Manager}
\label{sec:arena:manager}

The main responsibility of the battle manager is to pair two CRSs for a pairwise comparison, i.e., ``battle.'' The current matchmaking algorithm selects the two CRSs with the fewest recorded conversations, resolving ties by random selection, to ensure that all CRSs can be compared to each other as uniformly as possible. In the future, the algorithm may be improved to consider other factors, such as the performance of the CRSs in previous battles.

\vspace*{-.6\baselineskip}
\subsection{User Interface}
\label{sec:arena:ui}

\begin{figure}
    \shrink
    \centering
    \includegraphics[width=\columnwidth]{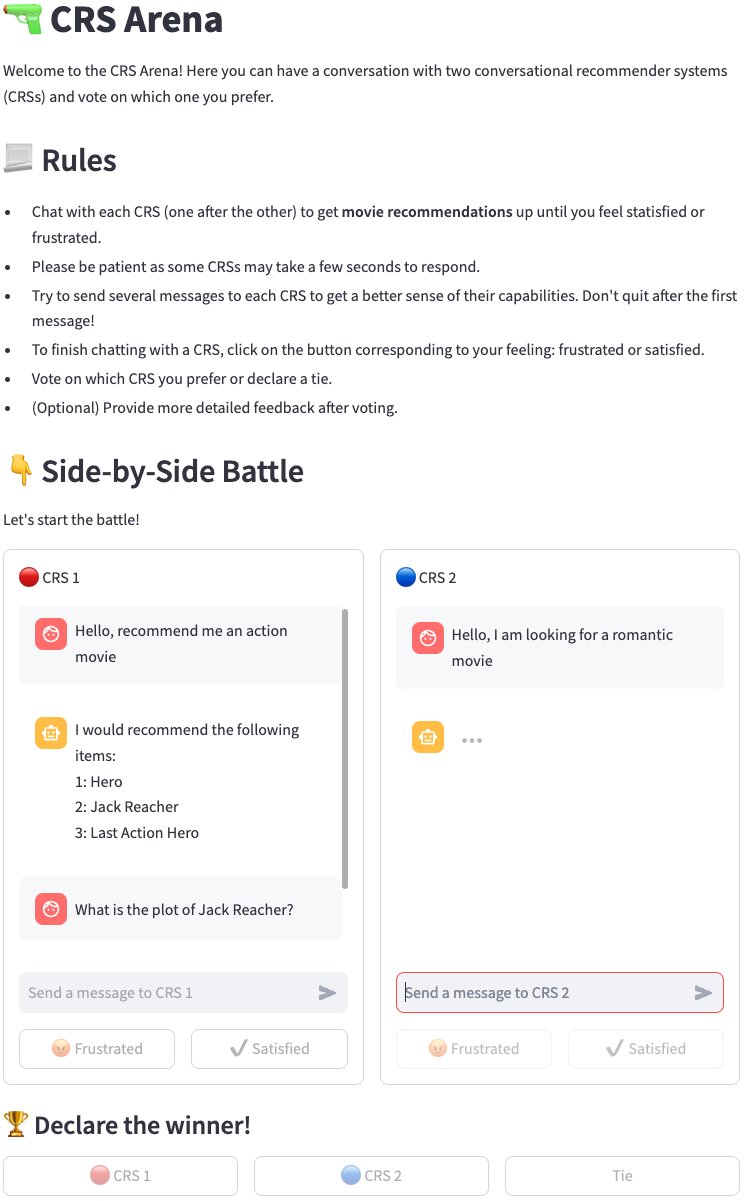}
    \shrink
    \caption{Screenshot of the CRS Arena.}
    \label{fig:arena}
    \shrink
\end{figure}

The user interface of CRS Arena is designed to be engaging and accessible to users with diverse backgrounds; see Fig.~\ref{fig:arena}. It is divided into five sections, as follows (from top to bottom):
\begin{enumerate}[leftmargin=0.5cm]
    \item \emph{Introduction} provides a succinct description of the platform's purpose.
    \item \emph{Rules} concisely outlines how sessions are conducted: users are expected to get movie recommendations by conversing with two anonymous CRSs sequentially, providing explicit feedback (frustration or satisfaction) for each. Once both conversations are completed, the user selects a winner or declares a draw.
    \item \emph{Side-by-side battles} is the central part of the frontend. It contains two chat interfaces associated with the two CRSs. A chat interface is composed of: (1) a chat window, with the conversation history and an input field to type messages and (2) two buttons to express satisfaction or frustration before ending the conversation. Note that once a conversation is ended, the user cannot go back to it.
    \item \emph{Vote} presents three buttons to choose a winner or declare a draw. After voting, the user has the option to provide additional feedback.
    \item \emph{Terms of service and contact} (not visible in Fig.~\ref{fig:arena}) explains that CRS Arena is a research platform that collects data and provides a point of contact for support.
\end{enumerate}
\vspace*{-.6\baselineskip}
\subsection{Implementation}
\label{sec:arena:imp}

CRS Arena is implemented as a web application using Streamlit,\footnote{\url{https://streamlit.io/}} a Python library for building interactive applications. At the beginning of each session, a user is assigned an identifier ensuring that data remains anonymous (unless the  user voluntarily discloses personal information during the conversations).

CRS Arena is publicly deployed on the HuggingFace Hub.\footnote{\url{https://huggingface.co/}} We observe that some models have high latency when interacting with users, which may affect their experience. It is likely due to high traffic and hardware limitations as the platform currently runs on 2 CPU cores and 16 GB of RAM. 

\section{Data and Analysis}
\label{sec:analysis}

CRS Arena has been publicly released in September 2024. After a period of approximately 10 days, we collected 254 conversations and 81 votes from users.
Note that we do not apply any filtering, e.g., based on quality, to preserve the raw and uncurated nature of the data.
Additionally, we conducted a study in a closed environment, with access restricted to selected crowd workers, to assess the robustness of CRS Arena across different user groups.
Specifically, we recruited workers from English-speaking countries on Prolific, with a 100\% approval rate and $\geq$ 1000 previous submissions, to ensure data quality. The data was collected within approximately 7 hours, with each battle taking around 9 minutes at a cost of roughly £1.33. Considering that some workers might behave to maximize their financial gain~\citep{Eickhoff:2011:CSDM}, which could lead to a limited number of interactions, we explicitly instructed them to interact with each CRS at least 5 times. Using this setup, we collected 220 conversations and 104 votes.
Overall, the number of conversations collected from the different CRSs is fairly uniform.\footnote{A detailed analysis including the conversation distribution is available at: \url{https://github.com/iai-group/crsarena-dial/blob/main/DataAnalysis.md}.}

\begin{table*}
    \centering
    \footnotesize
    \caption{Preliminary ranking of systems and level of satisfaction across different environments. R@10 shows the recall of each CRS as reported in~\citet{Wang:2023:EMNLP}. The rank position is shown in parentheses for Elo rating and R@10.}
    \shrink
    \label{tab:comparison}
    \begin{tabular}{l|c|c|c|c|c|c|c}
        \hline
        \multirow{2}{*}{\textbf{CRS}} & \multicolumn{2}{c|}{\textbf{Open crowdsourcing}} & \multicolumn{2}{c|}{\textbf{Closed crowdsourcing}} & \multicolumn{2}{c |}{\textbf{Closed \& open crowdsourcing}} & \multirow{2}{*}{\textbf{R@10 (rank)}} \\
        & \textbf{Elo (rank)} & \textbf{\% sat.} & \textbf{Elo (rank)} & \textbf{\% sat.} & \textbf{Elo (rank)} & \textbf{\% sat.} &  \\ \hline
        BARCOR\_OpenDialKG & 968 (7) & 17.2 & 1008 (4) & 11.5 & 988 (5) & 14.5 & 0.453 (3) \\
        BARCOR\_ReDial & 1052 (2)  & 30.4 & 1044 (3) & 29.2 & 1077 (3) & 29.8 & 0.170 (7) \\
        CRB-CRS\_ReDial & 930 (9) & 5.4 & 964 (6) & 11.5 & 930 (8) & 7.9 & --- \\
        ChatGPT\_OpenDialKG & \textbf{1056 (1)} & 50 & 1066 (2) & \textbf{54.2} & \textbf{1102 (1)} & \textbf{52.3} & \textbf{0.539} (1) \\
        ChatGPT\_ReDial & 1036 (3) & \textbf{58.6} & \textbf{1085 (1)} & 29.2 & \textbf{1102 (1)} & 45.3 & 0.174 (6) \\
        KBRD\_OpenDialKG & 966 (8) & 2.6 & 942 (9) & 0.0 & 920 (9) & 1.7 & 0.423 (4) \\
        KBRD\_ReDial & 1027 (4) & 8.6 & 985 (5) & 7.4 & 994 (4) & 8.1 & 0.169 (8) \\
        UniCRS\_OpenDialKG & 974 (6) & 9.5 & 953 (7) & 0.0 & 937 (7) & 4.8 & 0.513 (2) \\
        UniCRS\_ReDial & 991 (5) & 18.2 & 952 (8) & 3.7 & 950 (6) & 10.2 & 0.215 (5) \\ \hline
    \end{tabular}
\end{table*}

The evaluation of CRSs based on users' ratings is presented in Table~\ref{tab:comparison}. We observe a generally low level of user satisfaction in both environments, with ChatGPT\_OpenDialKG and ChatGPT\_ReDial achieving the highest average satisfaction rates of 52.1\% and 43.9\%, respectively.  Notably, even the best-performing CRSs can only satisfy users in approximately half of the cases.
Detailed user feedback indicates a few reasons for frustration, such as the lack of understanding of users' requests leading to irrelevant responses and recommendations, repetitive answers in some cases, lack of certain information, and latency issues.

Table~\ref{tab:comparison} also presents a preliminary ranking of CRSs based on the Elo rating~\citep{Elo:1967:ChessLife}, computed from the 185 pairwise judgments collected. For Elo computation, initial rating is set to 1000 and K-factor is 16.
We note that this preliminary ranking is consistent with the feeling of satisfaction expressed by users in the collected conversations, as indicated by a strong Spearman correlation of $\rho = 0.917$.
The R@10 column shows the recall of each CRS reported in~\citet{Wang:2023:EMNLP}. The Spearman's coefficient between R@10 and Elo ratings is $\rho = -0.238$, showing a negative ranking correlation. This shows that performing well on the recommendation aspect does not align well with user satisfaction, thus, highlighting the importance of evaluating CRSs in a holistic manner.

\begin{table}
    \centering
    \footnotesize
    \caption{Comparison between conversations collected in open and closed crowdsourcing environments.}
    \shrink
    \label{tab:statistics}
    \begin{tabular}{lcc}
        \hline
        \textbf{Statistics} & \textbf{ Open} & \textbf{Closed} \\
        \hline
        \#Utterances per dialogue & 8.13 & 11.15 \\
        \#Words per utterance & 11.01 & 11.59 \\
        Diversity (Distinct-2) & 0.527 & 0.538 \\
        \hline
    \end{tabular}
    \shrink
\end{table}

There is a strong correlation when comparing the open and closed crowdsourcing environments. 
Indeed, for Elo ratings, Pearson's and Spearman's correlation coefficients are $r = 0.763$ and $\rho = 0.700$, respectively, while for satisfaction, $r = 0.843$ and $\rho = 0.726$. Table~\ref{tab:statistics} compares high-level characteristics of conversations from both environments. Response diversity is measured using Distinct-2, following~\citet{Joko:2024:SIGIR}, with the same hyperparameters. Both environments show similar statistics, with the average number of utterances per dialogue being slightly longer in the closed one, as crowd workers were instructed to have at least 5 interactions. Overall, the results suggest that user feedback and conversation characteristics are similar in both open and closed setups, indicating the robustness of CRS Arena across different user groups. This is particularly important because using a closed crowdsourcing setup enables fast and scalable human evaluation of CRSs when a new system is released.

We release the resulting dataset, CRSArena-Dial,\footnote{\url{https://github.com/iai-group/crsarena-dial}} which represents a unique resource. Indeed, unlike existing dialogue corpora, it contains conversations with multiple CRSs and real users, in addition to pairwise comparisons between the systems. 
We acknowledge that the dataset is noisy due to the minimally regulated environment, where only high-level guidelines were provided. However, it remains highly representative of how regular users naturally express preferences and seek recommendations. We thus believe that the dataset is a valuable resource for evaluating CRSs and studying authentic user interactions with such systems.

\vspace*{-.6\baselineskip}
\section{Conclusion}
\label{sec:concl}

We introduced CRS Arena for benchmarking conversational recommender systems in a realistic setting. 
The platform anonymously collects conversations between CRSs and real users, along with feedback and pairwise CRS preferences.
We released CRSArena-Dial, a dataset comprising 474 conversations and their corresponding user feedback, collected in open and closed crowdsourced environments. Analysis of the data reveals an overall low level of satisfaction, highlighting that further research is needed to improve the quality of CRSs.
We also presented a preliminary ranking of nine CRSs based on Elo ratings, providing an early snapshot of a leaderboard.

As a resource built for the community, we rely on active contri\-butions---both in adding new CRSs and participating as users. This, in turn, would allow for creation and release of a larger and more diverse version of CRSArena-Dial and enable a more comprehensive evaluation of CRSs. As the community grows, we also plan to scale up the platform to be able to accommodate more users.

\begin{acks}
  We thank all the contributors of iEvaLM. This research was partially supported by the Norwegian Research Center for AI Innovation, NorwAI (309834, Research Council of Norway) and the Dutch Research Council (NWO) under the LESSEN project (NWA.1389.20.183).
\end{acks}
\shrink

\bibliographystyle{ACM-Reference-Format}
\bibliography{wsdm2025-arena.bib}


\begin{thebibliography}{20}

%%% the Web address.

\ifx \showCODEN    \undefined \def \showCODEN     #1{\unskip}     \fi
\ifx \showDOI      \undefined \def \showDOI       #1{#1}\fi
\ifx \showISBNx    \undefined \def \showISBNx     #1{\unskip}     \fi
\ifx \showISBNxiii \undefined \def \showISBNxiii  #1{\unskip}     \fi
\ifx \showISSN     \undefined \def \showISSN      #1{\unskip}     \fi
\ifx \showLCCN     \undefined \def \showLCCN      #1{\unskip}     \fi
\ifx \shownote     \undefined \def \shownote      #1{#1}          \fi
\ifx \showarticletitle \undefined \def \showarticletitle #1{#1}   \fi
\ifx \showURL      \undefined \def \showURL       {\relax}        \fi
% invisible to TeX
\providecommand\bibfield[2]{#2}
\providecommand\bibinfo[2]{#2}
\providecommand\natexlab[1]{#1}
\providecommand\showeprint[2][]{arXiv:#2}

\bibitem[Balog and Zhai(2024)]%
        {Balog:2024:FnTIR}
\bibfield{author}{\bibinfo{person}{Krisztian Balog} {and} \bibinfo{person}{ChengXiang Zhai}.} \bibinfo{year}{2024}\natexlab{}.
\newblock \showarticletitle{User Simulation for Evaluating Information Access Systems}.
\newblock \bibinfo{journal}{\emph{Found. Trends Inf. Retr.}} \bibinfo{volume}{18}, \bibinfo{number}{1-2} (\bibinfo{year}{2024}), \bibinfo{pages}{1--261}.
\newblock


\bibitem[Carterette et~al\mbox{.}(2008)]%
        {Carterette:2008:ECIR}
\bibfield{author}{\bibinfo{person}{Ben Carterette}, \bibinfo{person}{Paul~N. Bennett}, \bibinfo{person}{David~Maxwell Chickering}, {and} \bibinfo{person}{Susan~T. Dumais}.} \bibinfo{year}{2008}\natexlab{}.
\newblock \showarticletitle{Here or there: preference judgments for relevance}. In \bibinfo{booktitle}{\emph{Proc. of ECIR '08}}. \bibinfo{pages}{16--27}.
\newblock


\bibitem[Chen et~al\mbox{.}(2019)]%
        {Chen:2019:EMNLP}
\bibfield{author}{\bibinfo{person}{Qibin Chen}, \bibinfo{person}{Junyang Lin}, \bibinfo{person}{Yichang Zhang}, \bibinfo{person}{Ming Ding}, \bibinfo{person}{Yukuo Cen}, \bibinfo{person}{Hongxia Yang}, {and} \bibinfo{person}{Jie Tang}.} \bibinfo{year}{2019}\natexlab{}.
\newblock \showarticletitle{Towards Knowledge-Based Recommender Dialog System}. In \bibinfo{booktitle}{\emph{Proc. of EMNLP-IJCNLP '19}}. \bibinfo{pages}{1803--1813}.
\newblock


\bibitem[Chiang et~al\mbox{.}(2024)]%
        {Chiang:2024:ICML}
\bibfield{author}{\bibinfo{person}{Wei-Lin Chiang}, \bibinfo{person}{Lianmin Zheng}, \bibinfo{person}{Ying Sheng}, \bibinfo{person}{Anastasios~Nikolas Angelopoulos}, \bibinfo{person}{Tianle Li}, \bibinfo{person}{Dacheng Li}, \bibinfo{person}{Banghua Zhu}, \bibinfo{person}{Hao Zhang}, \bibinfo{person}{Michael Jordan}, \bibinfo{person}{Joseph~E. Gonzalez}, {and} \bibinfo{person}{Ion Stoica}.} \bibinfo{year}{2024}\natexlab{}.
\newblock \showarticletitle{Chatbot Arena: An Open Platform for Evaluating {LLM}s by Human Preference}. In \bibinfo{booktitle}{\emph{Proc. of ICML '24}}.
\newblock


\bibitem[Eickhoff and de~Vries(2011)]%
        {Eickhoff:2011:CSDM}
\bibfield{author}{\bibinfo{person}{Carsten Eickhoff} {and} \bibinfo{person}{Arjen~P. de Vries}.} \bibinfo{year}{2011}\natexlab{}.
\newblock \showarticletitle{How Crowdsourcable is Your Task?}. In \bibinfo{booktitle}{\emph{Proc. of CSDM '11}}. \bibinfo{pages}{11--14}.
\newblock


\bibitem[Elo(1967)]%
        {Elo:1967:ChessLife}
\bibfield{author}{\bibinfo{person}{Arpad~E Elo}.} \bibinfo{year}{1967}\natexlab{}.
\newblock \showarticletitle{The proposed uscf rating system, its development, theory, and applications}.
\newblock \bibinfo{journal}{\emph{Chess life}} \bibinfo{volume}{22}, \bibinfo{number}{8} (\bibinfo{year}{1967}), \bibinfo{pages}{242--247}.
\newblock


\bibitem[Gao et~al\mbox{.}(2021)]%
        {Gao:2021:AIOpen}
\bibfield{author}{\bibinfo{person}{Chongming Gao}, \bibinfo{person}{Wenqiang Lei}, \bibinfo{person}{Xiangnan He}, \bibinfo{person}{Maarten {de Rijke}}, {and} \bibinfo{person}{Tat-Seng Chua}.} \bibinfo{year}{2021}\natexlab{}.
\newblock \showarticletitle{Advances and challenges in conversational recommender systems: A survey}.
\newblock \bibinfo{journal}{\emph{AI Open}}  \bibinfo{volume}{2} (\bibinfo{year}{2021}), \bibinfo{pages}{100--126}.
\newblock


\bibitem[Jannach et~al\mbox{.}(2021)]%
        {Jannach:2021:CSUR}
\bibfield{author}{\bibinfo{person}{Dietmar Jannach}, \bibinfo{person}{Ahtsham Manzoor}, \bibinfo{person}{Wanling Cai}, {and} \bibinfo{person}{Li Chen}.} \bibinfo{year}{2021}\natexlab{}.
\newblock \showarticletitle{A Survey on Conversational Recommender Systems}.
\newblock \bibinfo{journal}{\emph{ACM Comput. Surv.}} \bibinfo{volume}{54}, \bibinfo{number}{5} (\bibinfo{year}{2021}), \bibinfo{pages}{1--36}.
\newblock


\bibitem[Jin et~al\mbox{.}(2024)]%
        {Jin:2024:TOIS}
\bibfield{author}{\bibinfo{person}{Yucheng Jin}, \bibinfo{person}{Li Chen}, \bibinfo{person}{Wanling Cai}, {and} \bibinfo{person}{Xianglin Zhao}.} \bibinfo{year}{2024}\natexlab{}.
\newblock \showarticletitle{CRS-Que: A User-centric Evaluation Framework for Conversational Recommender Systems}.
\newblock \bibinfo{journal}{\emph{ACM Trans. Recomm. Syst.}} \bibinfo{volume}{2}, \bibinfo{number}{1} (\bibinfo{year}{2024}).
\newblock


\bibitem[Joko et~al\mbox{.}(2024)]%
        {Joko:2024:SIGIR}
\bibfield{author}{\bibinfo{person}{Hideaki Joko}, \bibinfo{person}{Shubham Chatterjee}, \bibinfo{person}{Andrew Ramsay}, \bibinfo{person}{Arjen~P. de Vries}, \bibinfo{person}{Jeff Dalton}, {and} \bibinfo{person}{Faegheh Hasibi}.} \bibinfo{year}{2024}\natexlab{}.
\newblock \showarticletitle{Doing Personal LAPS: LLM-Augmented Dialogue Construction for Personalized Multi-Session Conversational Search}. In \bibinfo{booktitle}{\emph{Proc. of SIGIR '24}}. \bibinfo{pages}{796--806}.
\newblock


\bibitem[Kulikov et~al\mbox{.}(2019)]%
        {Kulikov:2019:ILNG}
\bibfield{author}{\bibinfo{person}{Ilia Kulikov}, \bibinfo{person}{Alexander Miller}, \bibinfo{person}{Kyunghyun Cho}, {and} \bibinfo{person}{Jason Weston}.} \bibinfo{year}{2019}\natexlab{}.
\newblock \showarticletitle{Importance of Search and Evaluation Strategies in Neural Dialogue Modeling}. In \bibinfo{booktitle}{\emph{Proc. of INLG '19'}}. \bibinfo{pages}{76--87}.
\newblock


\bibitem[Li et~al\mbox{.}(2018)]%
        {Li:2018:NIPS}
\bibfield{author}{\bibinfo{person}{Raymond Li}, \bibinfo{person}{Samira Kahou}, \bibinfo{person}{Hannes Schulz}, \bibinfo{person}{Vincent Michalski}, \bibinfo{person}{Laurent Charlin}, {and} \bibinfo{person}{Chris Pal}.} \bibinfo{year}{2018}\natexlab{}.
\newblock \showarticletitle{Towards Deep Conversational Recommendations}. In \bibinfo{booktitle}{\emph{Proc. of NIPS '18}}. \bibinfo{pages}{9748--9758}.
\newblock


\bibitem[Manzoor and Jannach(2021)]%
        {Manzoor:2021:RecSys}
\bibfield{author}{\bibinfo{person}{Ahtsham Manzoor} {and} \bibinfo{person}{Dietmar Jannach}.} \bibinfo{year}{2021}\natexlab{}.
\newblock \showarticletitle{Generation-based vs. Retrieval-based Conversational Recommendation: A User-Centric Comparison}. In \bibinfo{booktitle}{\emph{Proc. of RecSys '21}}. \bibinfo{pages}{515--520}.
\newblock


\bibitem[Manzoor and Jannach(2022a)]%
        {Manzoor:2022:KARS}
\bibfield{author}{\bibinfo{person}{Ahtsham Manzoor} {and} \bibinfo{person}{Dietmar Jannach}.} \bibinfo{year}{2022}\natexlab{a}.
\newblock \showarticletitle{INFACT: An Online Human Evaluation Framework for Conversational Recommendation}. In \bibinfo{booktitle}{\emph{Proc. of KARS '22}}. \bibinfo{pages}{6--11}.
\newblock


\bibitem[Manzoor and Jannach(2022b)]%
        {Manzoor:2022:InfSys}
\bibfield{author}{\bibinfo{person}{Ahtsham Manzoor} {and} \bibinfo{person}{Dietmar Jannach}.} \bibinfo{year}{2022}\natexlab{b}.
\newblock \showarticletitle{Towards retrieval-based conversational recommendation}.
\newblock \bibinfo{journal}{\emph{Inf. Sys.}}  \bibinfo{volume}{109} (\bibinfo{year}{2022}), \bibinfo{pages}{102083}.
\newblock


\bibitem[Moon et~al\mbox{.}(2019)]%
        {Moon:2019:ACL}
\bibfield{author}{\bibinfo{person}{Seungwhan Moon}, \bibinfo{person}{Pararth Shah}, \bibinfo{person}{Anuj Kumar}, {and} \bibinfo{person}{Rajen Subba}.} \bibinfo{year}{2019}\natexlab{}.
\newblock \showarticletitle{{O}pen{D}ial{KG}: Explainable Conversational Reasoning with Attention-based Walks over Knowledge Graphs}. In \bibinfo{booktitle}{\emph{Proc. of ACL '19'}}. \bibinfo{pages}{845--854}.
\newblock


\bibitem[Wang et~al\mbox{.}(2022a)]%
        {Wang:2022:arXiv}
\bibfield{author}{\bibinfo{person}{Ting-Chun Wang}, \bibinfo{person}{Shang-Yu Su}, {and} \bibinfo{person}{Yun-Nung Chen}.} \bibinfo{year}{2022}\natexlab{a}.
\newblock \bibinfo{title}{BARCOR: Towards A Unified Framework for Conversational Recommendation Systems}.
\newblock
\newblock
\showeprint[arxiv]{2203.14257}~[cs.CL]


\bibitem[Wang et~al\mbox{.}(2023)]%
        {Wang:2023:EMNLP}
\bibfield{author}{\bibinfo{person}{Xiaolei Wang}, \bibinfo{person}{Xinyu Tang}, \bibinfo{person}{Xin Zhao}, \bibinfo{person}{Jingyuan Wang}, {and} \bibinfo{person}{Ji-Rong Wen}.} \bibinfo{year}{2023}\natexlab{}.
\newblock \showarticletitle{Rethinking the Evaluation for Conversational Recommendation in the Era of Large Language Models}. In \bibinfo{booktitle}{\emph{Proc. of EMNLP '23}}. \bibinfo{pages}{10052--10065}.
\newblock


\bibitem[Wang et~al\mbox{.}(2022b)]%
        {Wang:2022:KDD}
\bibfield{author}{\bibinfo{person}{Xiaolei Wang}, \bibinfo{person}{Kun Zhou}, \bibinfo{person}{Ji-Rong Wen}, {and} \bibinfo{person}{Wayne~Xin Zhao}.} \bibinfo{year}{2022}\natexlab{b}.
\newblock \showarticletitle{Towards Unified Conversational Recommender Systems via Knowledge-Enhanced Prompt Learning}. In \bibinfo{booktitle}{\emph{Proc. of KDD '22}}. \bibinfo{pages}{1929--1937}.
\newblock


\bibitem[Zhou et~al\mbox{.}(2021)]%
        {Zhou:2021:ACL}
\bibfield{author}{\bibinfo{person}{Kun Zhou}, \bibinfo{person}{Xiaolei Wang}, \bibinfo{person}{Yuanhang Zhou}, \bibinfo{person}{Chenzhan Shang}, \bibinfo{person}{Yuan Cheng}, \bibinfo{person}{Wayne~Xin Zhao}, \bibinfo{person}{Yaliang Li}, {and} \bibinfo{person}{Ji-Rong Wen}.} \bibinfo{year}{2021}\natexlab{}.
\newblock \showarticletitle{{CRSL}ab: An Open-Source Toolkit for Building Conversational Recommender System}. In \bibinfo{booktitle}{\emph{Proc. of ACL-IJCNLP '21}}. \bibinfo{pages}{185--193}.
\newblock


\end{thebibliography}

\end{document}